# Explicitly correlated trial wave functions in Quantum Monte Carlo calculations of excited states of Be and Be$^-$

Luca Bertini[a*], Massimo Mella[a†], Dario Bressanini[b‡], and Gabriele Morosi[b§]

*a) Dipartimento di Chimica Fisica ed Elettrochimica, Universita` di Milano, via Golgi 19, 20133 Milano, Italy.*

*b) Dipartimento di Scienze Chimiche, Fisiche e Matematiche, Universita' dell'Insubria, via Lucini 3, 22100 Como, Italy*

---

[*] E-mail: bert@csrsrc.mi.cnr.it
[†] E-mail: Massimo.Mella@unimi.it
[‡] E-mail: Dario.Bressanini@uninsubria.it
[§] E-mail: Gabriele.Morosi@uninsubria.it

## Abstract

We present a new form of explicitly correlated wave function whose parameters are mainly linear, to circumvent the problem of the optimization of a large number of non-linear parameters usually encountered with basis sets of explicitly correlated wave functions. With this trial wave function we succeeded in minimizing the energy instead of the variance of the local energy, as is more common in quantum Monte Carlo methods. We applied this wave function to the calculation of the energies of Be $^3$P ($1s^22p^2$) and Be$^-$ $^4$S$^o$ ($1s^22p^3$) by variational and diffusion Monte Carlo methods. The results compare favorably with those obtained by different types of explicitly correlated trial wave functions already described in the literature. The energies obtained are improved with respect to the best variational ones found in literature, and within one standard deviation from the estimated non-relativistic limits.

# Introduction

The description of the electron correlation plays a central role in highly accurate quantum chemistry calculations. Mean-field methods give a qualitative description for many atomic and molecular systems, but in order to get quantitative results the instantaneous correlation between electrons must be taken into account. The most common way to include correlation is, starting from the Hartree-Fock picture, to approximate the exact wave function using MC-SCF or CI expansions. Unfortunately methods based on the orbital approximation converge very slowly to the non-relativistic limit. The reason is that these wave functions include the interelectronic distances only in an implicit form. Furthermore this implicit dependence is quadratic instead of linear, so the cusp conditions [1] of the exact wave functions are reproduced only for infinite expansions.

A very efficient and effective approach to accurately describe the local behavior of the wave function when two electrons collide is the explicit inclusion of the interelectronic distances into an approximate wave function. Hylleraas [2], Pekeris [3], James and Coolidge [4], and Kolos and Wolniewicz [5-7] showed how to obtain very accurate results for two electron systems by including the interelectronic distance into the wave function. An alternative possibility is the construction of many-particle permutational symmetry adapted functions in hyperspherical coordinates [8, 9]. Unfortunately it is not easy to generalize these methods to many-electron systems since the resulting integrals are extremely difficult to evaluate analytically. Beyond four electron systems, with at most two nuclei, the analytical approach becomes almost unfeasible [10, 11].

Instead of computing the integrals analytically, one could resort to a numerical method. The variational Monte Carlo (VMC) method [12, 13] is a very powerful numerical technique that estimates the energy, and all the desired properties, of a given trial wave function without any need to analytically compute the matrix elements. For this reason it poses no restriction on the functional form of the trial wave function, requiring only the evaluation of the wave function value, its gradient and its Laplacian, and these are easily computed. Using the VMC algorithm, essentially a stochastic numerical integration scheme, the expectation value of the energy for any form of the trial wave function can be estimated by averaging the local energy $\hat{H}\Psi_T(\mathbf{R})/\Psi_T(\mathbf{R})$ over an ensemble of configurations distributed as $\Psi_T^2$, sampled during a random walk in the configuration space using Metropolis [14] or Langevin algorithms [15]. The fluctuations of the local energy depend on the quality of the function $\Psi_T$, and they are zero if the exact wave function is used (zero variance principle). VMC can also be used to optimize the trial wave function $\Psi_T$, and we refer the reader to the literature for the technical details.

A popular and effective approach to building compact explicitly correlated wave functions is to multiply a determinantal wave function by a correlation factor, the most commonly used being a Jastrow factor [16]. The inclusion of the Jastrow factor does not allow the analytical evaluation of the integrals, so the use of VMC is mandatory. However, departing from the usual determinantal wave function form can be very fruitful [11], allowing an accurate and, at the same time, compact description of atomic and molecular systems. Very few terms are needed to reach a good accuracy, in comparison to more common wave function forms.

The recovery of the remaining correlation energy can be done using the diffusion Monte Carlo (DMC) method. Since this method is already well described in the literature, we refer the reader to the available reviews [12, 13]. We only recall here that in this method the exact, but unknown, wave function is interpreted as a probability density. In the fixed-node (FN) approximation [17] the nodal surfaces of the trial wave function $\Psi_T$ are used to partition the space and within each region the wave function can be safely interpreted as a probability density. It can be shown that the FN-DMC energies are an upper bound to the exact ground state energy.

This paper is part of an ongoing project in our laboratory to develop accurate and compact wave functions for few-electron systems.
In our previous works [11, 18, 19] we used linear expansions of explicitly correlated wave functions for calculations on the ground state of few-electron systems. In all cases good VMC energies were obtained, both in infinite nuclear mass approximation calculations and non-adiabatic calculations. In particular we used a linear expansion of explicitly correlated exponential functions to develop accurate wave functions for two test systems: the beryllium atom and the lithium hydride molecule in their ground state.

Here we present a new form of explicitly correlated wave function and we use VMC to extend the application of correlated trial wave function to excited states and five electron systems. Furthermore we use DMC to approximate the exact energies and compare them with the estimated non-relativistic limits.

We choose the Be $^3$P ($1s^22p^2$) and Be$^-$ $^4$S$^o$ ($1s^22p^3$) states which are involved in Beryllium electron affinity determination. We compare VMC and DMC energies and variances of the energy and examine the nodal properties of the trial wave functions comparing FN-DMC results with the best variational calculations and the non-relativistic limits estimated by Chung and coworkers [20, 21].

## Explicitly Correlated functional form

For an N electron atomic system we write an explicitly correlated trial wave function [11] as

$$\Psi = \hat{A}\{f(\mathbf{r})\phi\, g\, \Theta^N_{S,M_S}\} \tag{1}$$

In this equation $\hat{A}$ is the antisymmetrizer operator, $\phi$ is a function of all the electron-nucleus distances and g is a function of all the electron-electron distances called correlation factor. Both functions include variational parameters. $\Theta^N_{S,M_S}$ is an eigenfunction of the spin operators $\hat{S}^2$ and $\hat{S}_z$ of the correct spin multiplicity. The functions $\phi$ and g, being dependent only on interparticle distances, are rotationally invariant. This means that their product can describe only S states, with zero angular momentum. To describe higher angular momentum states, it is necessary to include a function f($\mathbf{r}$) with the correct rotational symmetry. f($\mathbf{r}$) is a function of the Cartesian electronic coordinates (x,y,z), but might include also the electron-nucleus distances [11]. This $\Psi$ function might be generalized including products of the interparticle distances, that is $\Psi$ is the two-body term of a many-body expansion of the wave function. It is possible to further generalize the wave function by taking linear combinations of such terms.

To assure a high quality wave function it is particularly important that the function $\Psi$ satisfy the cusp conditions [1], representing the behavior of the exact wave function at the coalescence of two particles. It is also important to take into account the asymptotic conditions [22], which represent the behavior when one of the particles go to infinity.

The first type of functional form we examined is generated assuming a Pade' factor $\exp[(ar+br^2)/(1+cr)]$ for the electron-nucleus part $\phi$ and a Jastrow factor $\exp[a'r/(1+c'r)]$ for the interelectronic part g.

$$\Psi = \hat{A}\left\{f(\mathbf{r})\exp\left[\sum_i \frac{ar_i + br_i^2}{1+cr_i}\right]\exp\left[\sum_{i<j}\frac{a'r_{ij}}{1+c'r_{ij}}\right]\Theta^{n_e}_{S,M_s}\right\} \tag{2}$$

In the following this wave function will be called Pade'-Jastrow. The Pade' factor is a good choice for the electron-nucleus part, because it is the best compromise between flexibility and small number of parameters. In fact this function goes as $e^{ar}$ for $r \to 0$ and $e^{(b/c)r}$ for $r \to \infty$. So with different exponents it can accommodate both the coalescence at the nucleus and the decay for large r. It is also important to point out that this factor can accurately describe both 1s and 2s orbitals as we have shown in our previous work [11].

The main problem with linear expansions of explicitly correlated trial wave functions is the huge number of non-linear parameters to optimize. In our previous work [11] for more sophisticated factors like Pade' or Jastrow we succeeded in optimizing trial wave functions including a maximum of two terms.

To overcome this problem we choose a second type of functional form, similar to the first one:

$$\Psi = \hat{A}\left\{ f(\mathbf{r})G(\mathbf{r})\exp\left[\sum_i \frac{ar_i + br_i^2}{1+cr_i}\right]\exp\left[\sum_{i<j}\frac{a'r_{ij}}{1+c'r_{ij}}\right]\Theta_{S,M_s}^{n_e}\right\} \quad (3)$$

We limit the expansion to a single term and so we have few non-linear parameters to optimize. However to add extra flexibility to the wave function we introduce a pre-exponential factor $G(\mathbf{r})$ written as a sum of powers of interparticle distances weighted by linear parameters:

$$G(\mathbf{r}) = \sum_p \sum_i g_p r_i^{n_p} + \sum_q \sum_{i<j} g_q r_{ij}^{n_q} + \sum_r \sum_i \sum_{j>i} g_r r_i^{n_r} r_{ij}^{o_r} + \sum_s \sum_i \sum_j g_s r_i^{n_s} r_j^{m_s} + \ldots \quad (4)$$

In the following we will call Eq. 3 pre-exponential wave function (prex).

Even if this kind of wave function allows us to reduce the effort for the optimization of the parameters, we recall that the CPU time needed to evaluate explicitly correlated trial wave functions is very large and proportional to the number of permutations generated by the antisymmetrizer. No matter the form we choose for explicitly correlated wave functions, they are limited to few electron systems.

In this paper we also compare Pade'-Jastrow and pre-exponential wave functions with a more standard form, widely used in QMC calculations [23, 24], that is the product of a determinantal function times the Schmidt-Moskowitz [25] correlation factor (SM):

$$\Psi = \sum_l C_l Det_l^\uparrow Det_l^\downarrow \exp\left[\sum_{i<j} U_{ij}(r_i, r_j, r_{ij})\right] \quad (5)$$

$Det^\uparrow$ and $Det^\downarrow$ are the determinants for $\alpha$ and $\beta$ electrons. The function U for atoms is given by

$$U_{ij} = \sum_k \Delta(m_k, n_k) c_k (\bar{r}_i^{m_k}\bar{r}_j^{n_k} + \bar{r}_j^{m_k}\bar{r}_i^{n_k})\bar{r}_{ij}^{o_k} \quad (6)$$

where $c_k$ are trial parameters and $\bar{r} = ar/(1+br)$.

The determinants are generated from ab initio calculations, in general SCF or MCSCF calculations, for a given basis set. Then the correlation factor is added, and its variational parameters optimized using VMC calculations.

## Optimization of the trial wave functions

Our previous work [11] showed that departing from the usual determinantal wave function form can be very fruitful, allowing to write very compact and at the same time very accurate wave functions. However it is computationally much more demanding and for this reason special care must be given to the design of an efficient way of generating and optimizing the trial wave function. These steps must be implemented in the most effective, fast and efficient way. The standard way to optimize a trial wave function using VMC is to minimize the variance of the local energy using a fixed sample of walkers; a method proposed by Frost [26] and Conroy [27] and described in detail by Umrigar, Wilson, and Wilkins [28] and by Mushinsky and Nightingale [29]. This has been proved to be numerically much more stable than the energy minimization. For our trial wave functions we have found very effective the minimization of the variance of the energy

$$\sigma^2(\hat{H}) = \langle \hat{H}^2 \rangle - \langle \hat{H} \rangle^2 \qquad (7)$$

or, even better, of the second moment with respect to an arbitrary parameter $E_R$, $\mu(E_R)$:

$$\mu(E_R) = \langle (\hat{H} - E_R)^2 \rangle = \sigma^2(\hat{H}) + (\langle \hat{H} \rangle - E_R)^2 \qquad (8)$$

where the parameter $E_R$ can be set equal to the exact energy of the system $E_0$. Both $\sigma(H)$ and $\mu(E_0)$ go to zero as $\Psi_T \to \Psi_0$, where $\Psi_0$ is the exact eigenfunction: their values for a given trial wave function $\Psi_T$ can be used to evaluate the quality of the trial wave function $\Psi_T$.

We used $\mu(E_R)$ as cost function for the optimization of both Pade'-Jastrow and SCF-MS wave functions.

As to Eq. 2, the optimization of the first term of the expansion is usually performed starting from a trial wave function with a reasonable electron-electron Jastrow factor, and with the electron-nucleus functions coming from some standard Slater orbital basis set, or from small basis sets optimized at the SCF level. As we showed in our previous work [11], it is possible to build a trial wave function as a linear expansion of n terms by adding an extra term to an optimized n-1 term wave function. This procedure worked well for two and three electron systems with simple exponential basis sets, but not for more sophisticated Pade' and Jastrow basis sets. For these reasons in this work we optimized only one term Pade'-Jastrow functions.

Let us now consider the case of the pre-exponential trial wave function. The function in Eq. 3 can be written as a linear combination:

$$\Psi = \sum_{l=1} g_l \Phi_l \qquad (9)$$

where the term $\Phi_l$ is given by

$$\Phi_l = \left[r_i^p....r_{ij}^m....\right]\Phi(\mathbf{p}) \tag{10}$$

$\Phi(\mathbf{p})$ is a single term Pade'-Jastrow function whose parameters **p** are optimized minimizing $\mu(E_R)$ in a preliminary step, and then are fixed during the optimization of the linear parameters.

As to the linear parameters, we succeeded in minimizing the energy instead of the variance of the local energy or the second moment $\mu(E_R)$, so we could choose the best linear parameters according to the observable we are interested in. The standard linear variational methods requires the solution of the secular problem, and so the calculation of the matrix elements

$$\begin{aligned} H_{ij} &= \int \Phi_i \hat{H} \Phi_j d\mathbf{R} \\ S_{ij} &= \int \Phi_i \Phi_j d\mathbf{R} \end{aligned} \tag{11}$$

These integrals are evaluated during a VMC simulation.

We show the main features of this pre-exponential trial wave function using as benchmarks the Be and Li ground states.

In table 1 we compare the energy for Be $^1S_0$ ground state obtained by the one term Pade'-Jastrow function with the value calculated by the function obtained adding a pre-exponential including all the electron-nucleus and electron-electron distances and their products, a total of 66 terms. The two values evidence a large gain of correlation energy for the addition of the pre-exponential factor and further improvement is obtained adding 20 more terms, that is the third and fourth powers of the electron-nucleus and electron-electron distances, to the linear expansion. So the pre-exponential factor adds flexibility to the wave function in a very efficient way.

To examine the relative efficiency of linear and non linear parameters in adding flexibility to the wave function, for the Li ground state (see table 2) we compare the energy of a 28-term pre-exponential function with the result of a 8-term expansion of explicitly correlated exponential functions, a simplified form of the Pade'-Jastrow wave function in which both the factor ϕ and the correlation factor g are in the form exp(cr). These two trial wave functions give a similar gain of correlation energy, but in the pre-exponential case there are 12 non-linear and 28 linear parameters, while the 8 term expansion of correlated exponential functions includes 48 non-linear and 8 linear parameters. In spite of the smaller number of parameters the pre-exponential function gives a better result, and its optimization process was much easier and faster.

We optimized the linear parameters also by minimization of $\mu(E_R)$. The calculated energy at VMC level is worse than the one obtained by minimization of the energy: this is obviously related to the different minima of the energy and $\mu(E_R)$.

## Results and discussion

For Be $^3$P ($1s^22p^2$) and Be$^-$ $^4$S$^o$ ($1s^22p^3$) we computed SCF and CASSCF trial wave functions using GAMESS with the Slater orbital basis sets reported in table 3. Each orbital was fitted with 6 Gaussian functions. We optimized the Schmidt-Moskowitz correlation factor of the SCF-SM functions and the non-linear parameters of the Pade'-Jastrow and pre-exponential functions minimizing the variance of the local energy.

Beside the VMC energies, we report the variance of the local energy, given by Eq. 7, estimated using VMC. DMC energies were obtained by a linear fit of the energy at three time steps ($\tau$=5, 3, and 1 mhartree$^{-1}$) and extrapolation to $\tau$=0 mhartree$^{-1}$.

We compare our results with the best variational energies obtained by Chung and coworkers [20, 21], who used linear expansions of Slater orbitals in the *L-S*-coupling scheme.

## Be $^3$P ($1s^22p^2$)

The calculations for the excited state Be $^3$P ($1s^22p^2$) were carried out with the following three trial wave functions:

1) a single determinantal function times a nine term Schmidt-Moskowitz correlation factor;
2) a one term Pade'-Jastrow function;
3) a pre-exponential function with 33 terms.

The spin eigenfunction used for the Pade'-Jastrow and pre-exponential function is $\Theta_{1,1}^4 = (\alpha\beta\alpha\alpha - \beta\alpha\alpha\alpha)$. The pre-factor f(**r**) that defines the state symmetry is

$$f(r) = x_3 y_4 - x_4 y_3 \qquad (12)$$

The results are reported in table 4.

As to the determinantal wave function, we used only a single determinant, as the energy lowers by only 0.003 hartree on going from the SCF to a CASSCF function for two electrons in an active space of two p shells (15 configurations, the highest weight of the first double excitation being equal to 0.04).
The explicit inclusion of the interelectronic distances in the wave function by the SM factor results in a large improvement of the quality of the wave function, as shown from the lowering of the energy (0.043 hartree) and the variance of the energy (almost four times smaller) on going from SCF to SCF-SM/VMC.

A further improvement of the trial wave function at VMC level is found using more sophisticated functional forms [11], like the Pade'-Jastrow and the pre-exponential ones. For the pre-exponential function we used a 33 term expansion, including all $r_i$, $r_i^2$, and the products $r_i r_{ij}$.

In particular in the case of the pre-exponential function we were able to optimize the linear parameters of G(**r**) minimizing the energy, not the variance of the local energy, obtaining an energy 0.9 mhartree higher than the best variational one. We also notice that this wave function is very compact with 45 (12 non-linear and 33 linear) variational parameters on the whole.

At DMC level already the SCF-SM wave function gives a lower energy than the best variational value, and 0.2 mhartree higher than the estimated non-relativistic limit (NRL). It means that the nodal surfaces of this function are fairly good, at variance with the SCF-SM trial wave function for the Be ground state, whose energy is 11 mhartree higher than the NRL [30]. This large nodal error is due to the strong contribution of the first double excitation in improving the quality of the nodal surfaces, because of the quasi-degeneracy of the 2s and 2p orbital.

DMC energies for Pade'-Jastrow and pre-exponential functions have the estimated NRL within one standard deviation, that is the nodal surfaces of these wave functions are correct and better than the SCF ones.

## Be$^-$ $^4$S$^o$ (1s$^2$ 2p$^3$)

The calculations for the excited state Be$^-$ $^4$S$^o$ were carried out with these four trial wave functions:

1) a single determinantal function times a nine term Moskowitz-Schmidt correlation factor;
2) a multideterminantal function times a nine term Moskowitz-Schmidt correlation factor;
3) a one term Pade'-Jastrow function;
4) a pre-exponential function G(**r**) with 61 terms.

The CASSCF wave function for three electrons in an active space of two p shells includes 20 configurations. The first two highest weights, relative to the first double and single excited configurations, are equal to 0.125 and 0.03 and indicate a more marked multiconfigurational character of the wave function.

As we have seen for Be $^3$P state, the gain in energy and the lowering of the variance of the energy between SCF and SCF-SM are very large, while between SCF-SM and CASSCF-SM they are an order of magnitude less.

The spin eigenfunction used for the Pade'-Jastrow and pre-exponential function is $\Theta^5_{3/2,3/2} = (\alpha\beta\alpha\alpha\alpha - \beta\alpha\alpha\alpha\alpha)$. The pre-factor f(r) that defines the state symmetry is

$$f(r) = x_3 y_4 z_5 + x_5 y_3 z_4 + x_4 y_5 z_3 - x_3 y_5 z_4 - x_5 y_4 z_3 - x_4 y_3 z_5 \qquad (13)$$

The results are reported is table 5.

For Be⁻ using the Pade' and pre-exponential functions we obtained better energies and variances of the energy than the SCF-SM and CASSCF-SM ones, as already seen in the Be $^3$P case. In particular, with the pre-exponential function with 61 linear parameters G(**r**) (all $r_i$, $r_i^2$ and the products $r_i r_{ij}$), the VMC energy is 0.9 mhartree higher than the best variational energy.

From DMC simulations we see clearly that the nodal surfaces don't change on going from the SCF to the CASSCF trial wave function and in both cases we have around 0.4 mhartree of nodal error.

Pade'-Jastrow and pre-exponential functions have better nodal surfaces and their DMC energies have the estimated NRL within one standard deviation.

## Conclusions

We have used explicitly correlated functional forms to improve the quality of the trial wave functions usually adopted to calculate the energy of a system. For the two excited states of Be and Be⁻ we obtained better non-relativistic energies with very compact trial wave functions compared to the best variational results.

Using a suitable pre-exponential factor we were able to improve the flexibility of the trial wave function without including too many non-linear parameters: this kind of trial wave function allowed us to minimize directly the energy instead of the variance of the local energy. As to the computational time, the optimization of the Be⁻ five electron trial wave function and the VMC calculation required around a week on a modern PC .It is not possible to compare this CPU time with calculations by correlated Gaussians as at present they are limited to four electron systems.

Our DMC energies are in good agreement with the estimated NRL obtained by Chung and coworkers [20, 21].

From our best values for Be $^3$P (-14.39547(5) hartree) and Be⁻ $^4$S$^o$ (-14.40620(6) hartree) we compute an electron affinity of 0.01073(8) hartree = 292(2) meV, within two standard deviations from the experimental value 295.49(25) meV [31] and the theoretical value 295.0(7) calculated by Hsu and Chung [20]. A significant comparison would require the reduction of the calculated standard deviation by one order of magnitude.

## Acknowledgements

CPU time for this work has been partially granted by the Centro CNR per lo studio delle relazioni tra struttura e reattività chimica, Milano.

Table 1. VMC results for Be ground state.

|  | VMC Energy (hartree) | % Correlation energy |
| --- | --- | --- |
| one term Pade'-Jastrow | -14.6528(2) | 84.57 |
| 66 term prex | -14.6633(3) | 95.70 |
| 86 term prex | -14.6651(2) | 97.60 |
| HF limit | -14.57302 |  |
| NR limit | -14.66735 |  |

Table 2. VMC results for Li ground state.

|  | VMC Energy (hartree) | % Correlation energy |
|---|---|---|
| 8 term exp. | -7.4775(2) | 98.29 |
| 28 term prex | -7.47770(8) | 99.20 |
| HF limit | -7.43274 |  |
| NR limit | -7.47806 |  |

Table 3. Basis sets for SCF and CASSCF calculations

| System | Be $^3$P | Be$^-$ $^4$S$^o$ |
|--------|----------|------------------|
| 1s | 5.7 | 5.7 |
| 1s | 4.2 | 4.2 |
| 2s | 4.3 | 4.3 |
| 2s | 2.4 | 2.4 |
| 2p | 1.65 | 1.65 |
| 2p | 0.76 | 0.76 |
| 2p |  | 0.376 |

Table 4. Be $^3$P ($1s^22p^2$) energies and VMC variances of the energy

| Method | Energy (hartree) | $\sigma_{VMC}$(H) |
|---|---|---|
| SCF | -14.3340 | 1.68(2) |
| SCF-SM/VMC | -14.3769(2) | 0.48(1) |
| Pade'/VMC | -14.3930(1) | 0.27(1) |
| 33 term prex /VMC | -14.3942(1) | 0.22(1) |
| Best variational | -14.3951086 | |
| SCF-SM/DMC | -14.39521(5) | |
| Pade'/DMC | -14.39541(7) | |
| 33 term prex /DMC | -14.39547(5) | |
| Estimated LNR | -14.3954404 | |

Table 5: Be $^4S^o$ ($1s^22p^3$) energies and VMC variances of the energy

| Method | Energy (hartree) | $\sigma_{VMC}$(H) |
|---|---|---|
| SCF | -14.326976 | 1.68(1) |
| CASSCF | -14.334010 | 1.68(1) |
| SCF-SM/VMC | -14.3769(2) | 0.48(1) |
| CASSCF-/VMC | -14.3836(1) | 0.48(1) |
| Pade'/VMC | -14.4031(2) | 0.29(1) |
| 61 term prex/VMC | -14.4051(2) | 0.21(1) |
| SCF-SM/DMC | -14.40594(8) | |
| CASSCF-SM/DMC | -14.40597(7) | |
| Best variational | -14.4060320 | |
| Pade'/DMC | -14.40620(4) | |
| 61 term prex /DMC | -14.40620(6) | |
| Estimated LNR | -14.406282(26) | |